\documentclass[letterpaper,twocolumn,superscriptaddress,floatfix]{revtex4}
\usepackage{inputenc}
\usepackage{bm}
\usepackage{multirow,amssymb,amsbsy,amsmath}
\usepackage{graphicx}
\usepackage{float}
\usepackage{verbatim}
\makeatletter
\usepackage{pifont}
\usepackage{upgreek}
\usepackage{color}
\usepackage{indentfirst}
\usepackage{soul}

\newcommand{\Rmnum}[1]{\expandafter\@slowromancap\romannumeral #1@}
\usepackage{caption}
\makeatother
\makeatletter %使\section中的内容左对齐
\renewcommand{\section}{\@startsection{section}{1}{0mm}
  {-\baselineskip}{0.5\baselineskip}{\bf\leftline}}
\makeatother

\begin{document}
\title{Reflective Dielectric Cavity Enhanced Emission from Hexagonal Boron Nitride Spin Defect Arrays}

\affiliation{CAS Key Laboratory of Quantum Information, University of Science and Technology of China, Hefei, Anhui 230026, China}
\affiliation{CAS Center For Excellence in Quantum Information and Quantum Physics,
University of Science and Technology of China, Hefei, Anhui 230026, China}
\affiliation{Hefei National Laboratory, University of Science and Technology of China, Hefei, Anhui 230088, China}

\author{Xiao-Dong Zeng}
\affiliation{CAS Key Laboratory of Quantum Information, University of Science and Technology of China, Hefei, Anhui 230026, China}
\affiliation{CAS Center For Excellence in Quantum Information and Quantum Physics,
University of Science and Technology of China, Hefei, Anhui 230026, China}

\author{Yuan-Ze Yang}
\affiliation{CAS Key Laboratory of Quantum Information, University of Science and Technology of China, Hefei, Anhui 230026, China}
\affiliation{CAS Center For Excellence in Quantum Information and Quantum Physics,
University of Science and Technology of China, Hefei, Anhui 230026, China}

\author{Nai-Jie Guo}
\affiliation{CAS Key Laboratory of Quantum Information, University of Science and Technology of China, Hefei, Anhui 230026, China}
\affiliation{CAS Center For Excellence in Quantum Information and Quantum Physics,
University of Science and Technology of China, Hefei, Anhui 230026, China}

\author{Zhi-Peng Li}
\affiliation{CAS Key Laboratory of Quantum Information, University of Science and Technology of China, Hefei, Anhui 230026, China}
\affiliation{CAS Center For Excellence in Quantum Information and Quantum Physics,
University of Science and Technology of China, Hefei, Anhui 230026, China}

\author{Zhao-An Wang}
\affiliation{CAS Key Laboratory of Quantum Information, University of Science and Technology of China, Hefei, Anhui 230026, China}
\affiliation{CAS Center For Excellence in Quantum Information and Quantum Physics,
University of Science and Technology of China, Hefei, Anhui 230026, China}

\author{Lin-Ke Xie}
\affiliation{CAS Key Laboratory of Quantum Information, University of Science and Technology of China, Hefei, Anhui 230026, China}
\affiliation{CAS Center For Excellence in Quantum Information and Quantum Physics,
University of Science and Technology of China, Hefei, Anhui 230026, China}

\author{Shang Yu}
\affiliation{CAS Key Laboratory of Quantum Information, University of Science and Technology of China, Hefei, Anhui 230026, China}
\affiliation{CAS Center For Excellence in Quantum Information and Quantum Physics,
University of Science and Technology of China, Hefei, Anhui 230026, China}

\author{Yu Meng}
\affiliation{CAS Key Laboratory of Quantum Information, University of Science and Technology of China, Hefei, Anhui 230026, China}
\affiliation{CAS Center For Excellence in Quantum Information and Quantum Physics,
University of Science and Technology of China, Hefei, Anhui 230026, China}

\author{Qiang Li}
\affiliation{CAS Key Laboratory of Quantum Information, University of Science and Technology of China, Hefei, Anhui 230026, China}
\affiliation{CAS Center For Excellence in Quantum Information and Quantum Physics,
University of Science and Technology of China, Hefei, Anhui 230026, China}

\author{Jin-Shi Xu}
\affiliation{CAS Key Laboratory of Quantum Information, University of Science and Technology of China, Hefei, Anhui 230026, China}
\affiliation{CAS Center For Excellence in Quantum Information and Quantum Physics,
University of Science and Technology of China, Hefei, Anhui 230026, China}
\affiliation{Hefei National Laboratory, University of Science and Technology of China, Hefei, Anhui 230088, China}
\author{Wei Liu}
\altaffiliation{Email: lw691225@ustc.edu.cn}
\affiliation{CAS Key Laboratory of Quantum Information, University of Science and Technology of China, Hefei, Anhui 230026, China}
\affiliation{CAS Center For Excellence in Quantum Information and Quantum Physics,
University of Science and Technology of China, Hefei, Anhui 230026, China}

\author{Yi-Tao Wang}
\altaffiliation{Email: yitao@ustc.edu.cn}
\affiliation{CAS Key Laboratory of Quantum Information, University of Science and Technology of China, Hefei, Anhui 230026, China}
\affiliation{CAS Center For Excellence in Quantum Information and Quantum Physics,
University of Science and Technology of China, Hefei, Anhui 230026, China}

\author{Jian-Shun Tang}
\altaffiliation{Email: tjs@ustc.edu.cn}
\affiliation{CAS Key Laboratory of Quantum Information, University of Science and Technology of China, Hefei, Anhui 230026, China}
\affiliation{CAS Center For Excellence in Quantum Information and Quantum Physics,
University of Science and Technology of China, Hefei, Anhui 230026, China}
\affiliation{Hefei National Laboratory, University of Science and Technology of China, Hefei, Anhui 230088, China}
\author{Chuan-Feng Li}
\altaffiliation{Email: cfli@ustc.edu.cn}
\affiliation{CAS Key Laboratory of Quantum Information, University of Science and Technology of China, Hefei, Anhui 230026, China}
\affiliation{CAS Center For Excellence in Quantum Information and Quantum Physics,
University of Science and Technology of China, Hefei, Anhui 230026, China}
\affiliation{Hefei National Laboratory, University of Science and Technology of China, Hefei, Anhui 230088, China}
\author{Guang-Can Guo}
\affiliation{CAS Key Laboratory of Quantum Information, University of Science and Technology of China, Hefei, Anhui 230026, China}
\affiliation{CAS Center For Excellence in Quantum Information and Quantum Physics,
University of Science and Technology of China, Hefei, Anhui 230026, China}
\affiliation{Hefei National Laboratory, University of Science and Technology of China, Hefei, Anhui 230088, China}

\begin{abstract}
Among the various kinds of spin defects in hBN, the negatively charged boron vacancy ($\rm V_B^-$) spin defect that can be deterministically generated is undoubtedly a potential candidate for quantum sensing, but its low quantum efficiency restricts its %use in
practical applications. Here, we demonstrate a robust enhancement structure with advantages including easy on-chip integration, convenient processing, low cost and suitable broad-spectrum enhancement for $\rm V_B^-$ defects. %Improved photoluminescence (PL) intensity and optically detected magnetic resonance (ODMR) contrast of $\rm V_B^-$ defect arrays. 
In the experiment, we used a metal reflective layer under the hBN flakes, filled with a transition dielectric layer in the middle, and adjusted the thickness of the dielectric layer to achieve the best coupling between the reflective dielectric cavity and the hBN spin defect. Using a reflective dielectric cavity, we achieved a PL enhancement of approximately 7-fold, and the corresponding ODMR contrast achieved  18\%. Additionally, the oxide layer of the reflective dielectric cavity can be used as an integrated material for micro-nano photonic devices for secondary processing, which means that it can be combined with other enhancement structures to achieve stronger enhancement. This work has guiding significance for realizing the on-chip integration of spin defects in two-dimensional materials.

\end{abstract}

\maketitle
\date{\today}
\section*{Introduction}

Solid-state optically addressable spin defects are highly sought-after quantum systems for realizing quantum information processing \cite{Sherson2006,Childress2013,Atatre2018,Awschalom2013,Togan2010}. The well-known systems are color centers in diamond \cite{Maze2008,Jelezko2006}, silicon carbide \cite{Koehl2011,Li2021} and two-dimensional (2D) materials  \cite{Caldwell Photonics-with-he,Li Carbon-defect-in-WSe2}. 2D materials are layered structures, whose layers are connected by van
der Waals (vd$\rm W$) forces. This quantum confinement effect perpendicular to the 2D plane results in many unique properties that are significantly different from those in bulk materials. Hexagonal boron nitride (hBN) is a wide-bandgap ($\sim$ 6 eV) 2D material, possessing good transferability and easy integration \cite{hBN is bandgap,Polymer transfer te}. Recent studies have shown the existence of color centers in hBN capable of spin initialization, readout, and coherent manipulation at room temperature \cite{Mendelson2021,Mende,Gottscholl2021,Photoluminescence pho,Gottscholl2020,Guo2021}.

\begin{figure*}[ht]
    \center
    \includegraphics[width=1\linewidth]{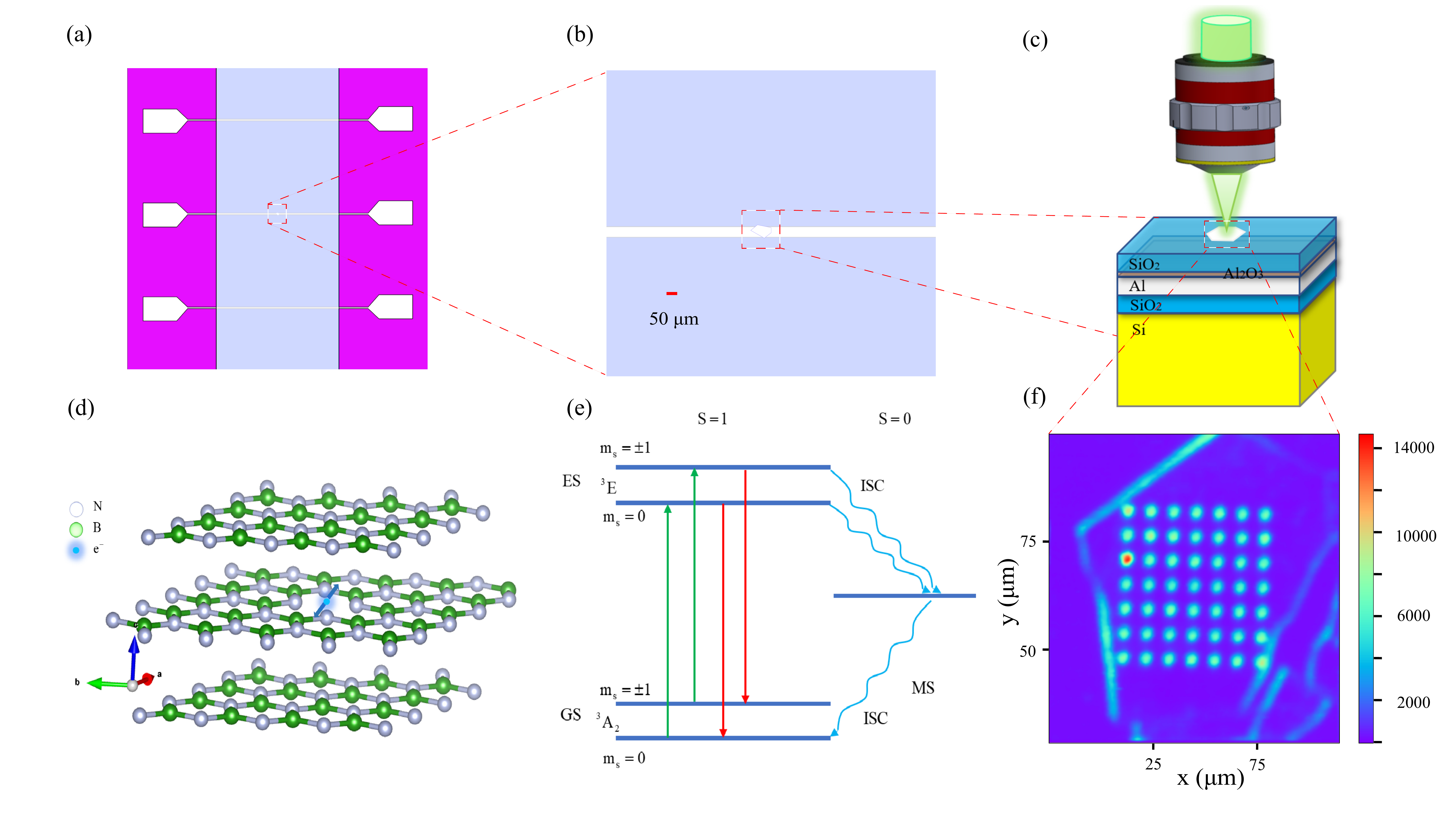}
    \caption{Reflective dielectric cavity structure and $\rm V_B^-$ spin defect energy level structure. (a) Schematic diagram of the coplanar waveguide substrate of the dielectric cavity, obtained from the processing of the $\rm SiO_2$/Si substrate, and the sample is transferred to the top of the waveguide. (b) The sample is located above the position of the waveguide enlarged, with a waveguide width of 50-$\rm \mu$m. (c) The hierarchy of the reflective dielectric cavity, from bottom to top, including 500-$\rm \mu$m Si layer, 285-nm $\rm SiO_2$ layer, 115-nm Al layer, 5-nm $\rm Al_2O_3$ layer, and $\rm SiO_2$ layer as an isolation medium. (d) Schematic of a $\rm V_B^-$ defect in hBN, where the boron atom is replaced by an electron. (e) Energy level structure diagram of the $\rm V_B^-$ spin defect, whose zero-field splitting (ZFS) of the ground state is approximately 3.5 GHz. (f) The hBN on the reflective dielectric cavity substrate was implanted with helium ions using HIM to generate a photoluminescence array of $\rm V_B^-$ spin defects. The excitation light was a 532-nm laser with a power of 4 mW.}
    \label{Figure1}
\end{figure*}

 The most studied spin defect in hBN is the negatively charged boron vacancy ($\rm V_B^-$) defect, which can exist stably at room temperature \cite{Stern2022,Liu2021} and can be deterministically generated in various ways, including neutron irradiation \cite{Gottscholl2020}, electron beam irradiation \cite{Murzakhanov2021}, focused ion beam (FIB) irradiation \cite{G2022,Kianinia2020}, laser ablation \cite{Gao2021,yang2022}, etc. Experiments have shown that the $\rm V_B^-$ defect a packet spectrum peaked at approximately 810 nm and a  nonobvious zero phonon line (ZPL) even at low temperature \cite{Gao2021}. Recently, some researchers have shown that the ZPL of $\rm V_B^-$ is at 773 nm through Purcell enhancement from the photonic crystal cavity \cite{Qian2022}, which is consistent with the theoretical results \cite{Ab2020,Photoluminescence1}. 
 %At present, there has been a deep  understanding of the structure of $\rm V_B^-$ spin defect \cite{Creation of Negatively,Edge effects on optically det}.
 The electron spin orientation of the $\rm V_B^-$ defect is out-of-plane relative to the lattice of hBN, whose ground state and excited state are triplet states. Previous works indicate that $\rm V_B^-$ has a ground-state splitting energy of $\sim$ 3.5 GHz \cite{Color centers in hexagonal boron} and an excited-state splitting energy of $\sim$ 2.1 GHz \cite{Yu2022,Mu2022,Baber2022,Mathur2022}. 
 Experiments have shown that hBN has great potential in sensing the  measurement of temperature \cite{Temperature-dependent en,Spin defects in hBN as promising}, stress \cite{Strain2022} and magnetic fields \cite{Quantum microscopy with van,Wide Field Imaging of van der Waals Ferromagnet}. 
 %However, the $\rm V_B^-$ has poor quantum efficiency, which will limit the sensitivity to external fields and hinder to finding the single $\rm V_B^-$ spin
 However, the poor quantum efficiency of $\rm V_B^-$ limits the sensitivity to external fields and hinders finding the single $\rm V_B^-$ spin \cite{Photoluminescence1}.

 To enhance the photoluminescence (PL) intensity of $\rm V_B^-$ defects, many methods have been applied, such as photonic crystal cavities \cite{Qian2022}, plasmonic cavities \cite{Mendelson2022,Gao2021_1}, bullseye cavities \cite{Coupling spin}, and strain engineering \cite{for strain sensing}. 
 Photonic crystal cavities are suitable for narrow-band, local enhancement, and have significant advantages in on-chip applications, but fabrication of photonic crystal cavities requires complex technology. In addition, there is a coupling problem between the color centers and the photonic crystal cavity \cite{Acs2020}. The enhancement effect of metal plasmons is remarkable and the experimental feasibility is high, but the strong coupling interaction between metals and 2D materials can seriously interfere with the PL spectra of 2D materials, making it difficult to analyze the natural properties of 2D materials \cite{Single type}.  Furthermore, some of these methods have limitations considering the strength of microwave radiation, resulting in low optically detected magnetic resonance (ODMR) contrast.

 In this paper, a reflective dielectric cavity (RDC) with convenient processing technology and strong scalability is used to enhance the PL intensity of $\rm V_B^-$ spin defects in hBN. 
 The RDC consisting of microwave coplanar waveguides as a metal reflector and $\rm SiO_2$ as a dielectric layer has been used for Raman and PL enhancement of other 2D materials,  fluorescent nanodiamonds, and quantum dot reinforcement \cite{Single type,Manipulating the distribution,Bright Quantum Dot Single-Photon}.
 Combining $\rm V_B^-$ defects with this structure, not only are the fluorescence counts enhanced 7 times but also the ODMR contrast reaches 18 $\rm \%$.
 This structure does not require complex micro-nano fabrication, and it is scalable and effective. In addition, the structure can be used to search for a single spin color center combined with other enhancement structures.

\begin{figure*}[ht]
    \centering
    \includegraphics[width=0.9\linewidth]{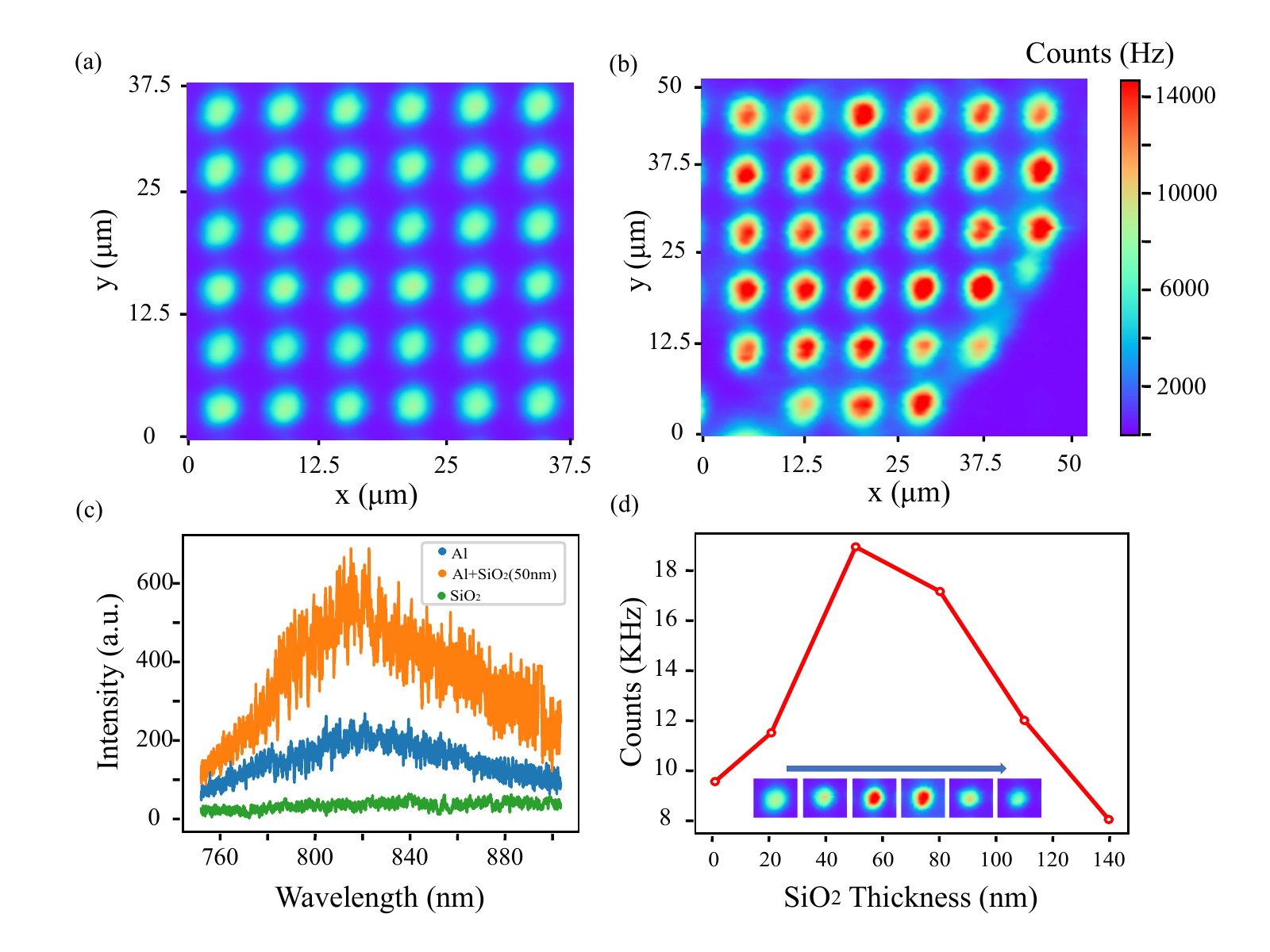}
    \caption{Optical properties of reflecting dielectric enhancement. (a) PL image of the spin defect array generated by helium ion implantation on an Al coplanar waveguide without a dielectric layer. (b) Spin defect array PL image on a 50-nm thick RDC. (c) The relative spectral intensities of different substrates under a 4 mW 532 nm laser. The measurement range is 750-900 nm. (d) The PL counts of the reflective dielectric cavity with different thick oxide layers, and the best enhancement effect was shown at 50 nm.}
    \label{Figure2}
\end{figure*}

\section*{Results}
As shown in Figure 1(a), we made aluminum (Al) electrodes as metal reflectors on the $\rm SiO_2/Si$ substrate and coated $\rm SiO_2$ as a dielectric layer. 
%in order to locate the sample in the reflective dielectric cavity with higher probability.
Subsequently, we transferred the flake hBN to the prepared
RDC, %and use optical microscope to observe the sample position. The selected hBN sample is located in the reflective dielectric cavity, 
and we deterministically generated the $\rm V_B^-$ spin defect array using a helium ion microscope (HIM) with an  implantation dose of $1 \times 10^{17}$ ions/$\rm cm^2$ and an implantation energy of 30 KeV. We scanned the PL map of $\rm V_B^-$ spin defects using a home-made confocal microscope as shown in Figure 1(f) (see Methods for details). Figure 1(c) shows the basic structure of the RDC, including the  $\rm SiO_2$/Si substrate, 115-nm thick aluminum (Al) coplanar waveguide, 5-nm thick $\rm Al_2O_3$ and dielectric isolation $\rm SiO_2$ layer from the bottom to the top. 
As shown in Figure 1(d), the $\rm V_B^-$ spin defect structure contains a boron atom bombarded out of the lattice and a captured electron. Figure 1(e) displays a schematic diagram of the energy level structure of $\rm V_B^-$ defects. 
\vspace{0cm}

\vspace{0cm}

\begin{figure*}[ht]
    \centering
    \includegraphics[width=1\linewidth]{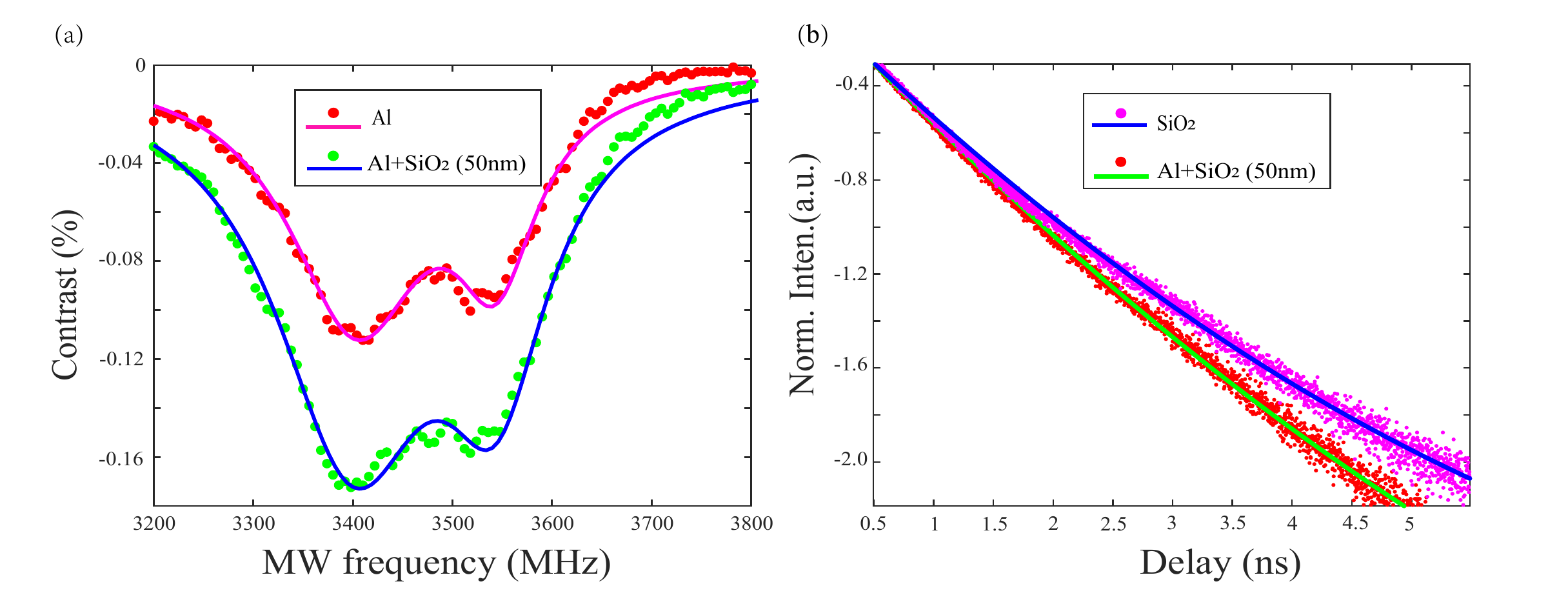}
    \caption{(a) The ODMR fitted data measured on the 0-nm RDC and 50-nm RDC. (b) The fluorescence lifetime on Si substrate and 50-nm RDC was measured. The solid lines are fitting results using $y=a*exp(-(t-b)/\tau)+c$, and the lifetimes are 0.825 ns and 0.787 ns, respectively, showing a weak Purcell effect.}
    \label{Figure 3}
\end{figure*}

Figure 2 (a) and (b) show the PL images of RDCs with 0-nm and 50-nm $\rm SiO_2$ dielectric layers (hereafter referred to as 0-nm RDC and 50-nm RDC, respectively) and the latter is observably brighter. Figure 2(c) shows the spectra of $\rm V_B^-$ spin defects on the $\rm SiO_2$/Si substrate, 0-nm RDC substrate and 50-nm RDC substrate. The spectral peak of $\rm V_B^-$ defects on 50-nm RDC is $\sim$ 10 times higher  than that on $\rm SiO_2$/Si. Then we explored the influence of the thickness of the dielectric layer on the enhancement effect, and the results are shown in Figure 2(d). With the increment of the thickness of the dielectric layer, the enhancement effect first increased and then decreased. The results show that the enhancement effect is optimized when the thickness of the dielectric layer is 50 nm, and the fluorescence counts are enhanced approximately 7 times compared with that of the $\rm SiO_2$/Si substrate.

Next, we measured the spin properties of $\rm V_B^-$ defects on different substrates.  Using the reflective metal layer as a microwave coplanar waveguide, we can perform ODMR measurements without additional microwave radiation structures. As shown in Figure 3(a), the ODMR spectra of 0-nm RDC and 50-nm RDC both have high contrast, and for the latter, the measured ODMR contrast can reach $\rm 18\%$, indicating that the RDC structure has a good effect on ODMR enhancement. Furthermore, the fluorescence lifetimes of $\rm V_B^-$ defects on different substrates have also been measured, because the Purcell enhancement by the plasmonic effect is always accompanied by a decrement of the fluorescence lifetime. Figure 3(b) shows the fluorescence lifetime of the $\rm V_B^-$ emitters on 50-nm RDC and SiO2/Si substrates. The fitted fluorescence lifetimes are 0.787 ns for the 50-nm RDC substrate and 0.825 ns for the $\rm SiO_2$/Si substrate. The lifetime of $\rm V_B^-$ emitters on RDC is only slightly shorter than that of $\rm V_B^-$ emitters on $\rm SiO_2$/Si, indicating that the Purcell enhancement effect is weak and not the major effect in our experiment. 
%The main effect is to enhance the  reflection and thus improve the collection efficiency. 

\begin{figure*}[ht]
    \centering
    \includegraphics[width=0.8\linewidth]{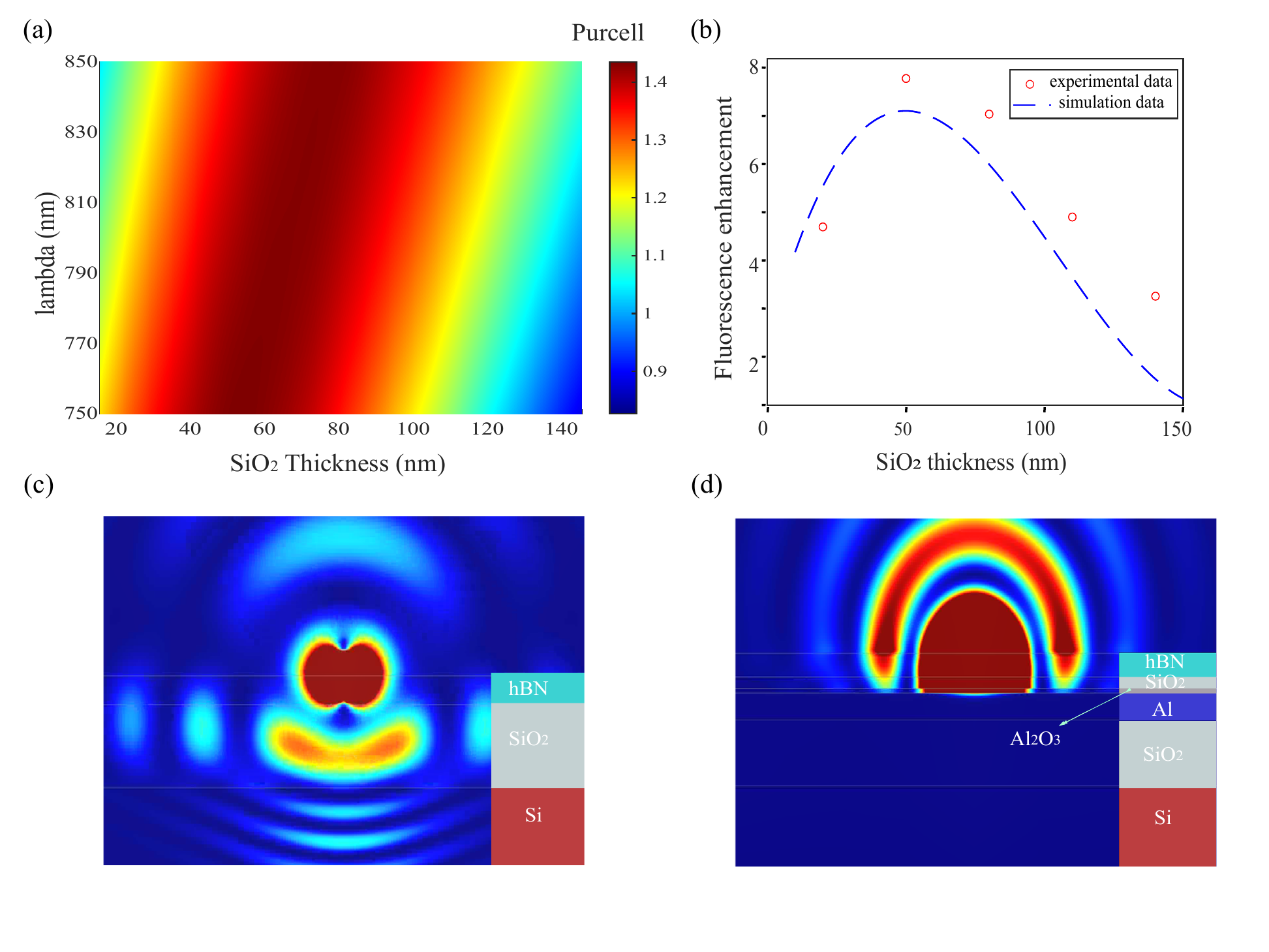}
    \caption{(a) The effect of the dielectric layer on the Purcell effect of the dielectric cavity under a wide spectrum dipole light source. We considered the oxidation effect of the Al coplanar waveguide, and set the oxide layer to 5 nm of $\rm Al_2O_3$. The results show that the dielectric layer of 61 nm has the strongest Purcell effect at wavelengths of  750-900 nm. (b) The experimental and simulated PL enhancement factors of 50-nm RDC relative to the $\rm SiO_2$/Si substrate. Considering the Purcell effect and the enhancement factor obtained from the simulation of the collection efficiency at a specific solid angle above the actual experimental sample, the simulation is in good agreement with the experiment. The simulation shows that the enhancement effect of the reflective medium cavity is mainly reflected in the improvement of the collection efficiency. (c) The propagation loss time slice of the dipole light source on the $\rm SiO_2$/Si substrate can be visually seen to be significantly dissipated. (d) The propagation loss time slice of the dipole light source on the reflective dielectric cavity of the 50-nm -thick medium layer shows that the loss becomes weaker and the directionality in the vertical upward direction becomes better, which is the main reason for the enhancement.}
    \label{Figure4}
\end{figure*}

To gain a deeper understanding of the physical mechanism of the enhanced RDC substrate, we performed a numerical simulation of the RDC structure.
Figure 4(a) shows the simulation results of the Purcell enhancement factor of a  broadband dipole source in the reflective dielectric cavity model. The Purcell factor first increases and then decreases with the increment of the thickness of the dielectric layer, and the extreme value is approximately  61 nm. RDC has a broad-spectrum enhancement effect. From Figure 4(a), we find that the RDC has a good enhancement for the entire wave packet of $\rm V_B^-$. The comparison between simulated (blue dotted line) and experimental (red circle) fluorescence enhancements is shown in Figure 4(b). The experimental data basically agree with the simulated results, and we speculate that the main error may come from the inaccuracy of the refractive index of the dielectric layer and the incomprehensive approximation of the environment.
%The dotted line is the simulation data corresponding to the enhancement rate of the reflective dielectric cavity compared to the $\rm SiO_2$/Si substrate, while the red circle is the experimental data, which is 7.6 times enhanced. 
Figures 4(c) and (d) are captured from the field propagation and distribution of the movie monitor in the simulation. The dissipation of spin defect luminescence on the $\rm SiO_2$/Si substrate is much greater than that of the reflection dielectric cavity. The existence of a reflective metal layer reduces the loss, and results in far-field collection efficiency due to the better orientation, which provides an intuitive explanation of the reflection dielectric cavity enhancement.

\section*{Discussions}
We explored the fluorescence enhancement effect and physical mechanism of the reflective dielectric cavity on the $\rm V_B^-$ defect array from the point of view of experiments and simulations. Enlarging the area of the coplanar waveguide can enhance the defects on a larger scale and be compatible with other structures. The dielectric layer can be etched as the Bragg circular resonator for hybrid enhancement \cite{hybrid circular Bragg resonator}. It is easy to extend this enhancement structure to enhance other 2D-materials using reflective dielectric cavities. For the spin defect array produced by HIM, some points appear inhomogeneous, which we attribute to the edge effect, stress or dangling bonds or other factors. The overall uniformity of the array produced by the HIM is good. Effects such as spectral drift also exist in many enhanced structures, limiting further applications, while the spectral features of the spin-defect ensemble on hBN coupled to a reflective dielectric cavity can be well preserved.

\section*{Conclusion}
Our experimental results show the enhancement of the fluorescence and ODMR contrast of the spin defect array by a reflective dielectric cavity. We explored the optimal dielectric cavity thickness and further explored the primary and secondary relationships of the enhancement related factors in combination with theoretical simulations. Since the $\rm V_B^-$ defect PL spectrum is a packet at room temperature, the reflective dielectric cavity has a good correspondence with the $\rm V_B^-$ spectrum range. This takes advantage of little enhancement to the background fluorescence, which is directly reflected in the improvement of ODMR contrast. From the simulation results, different from many enhancement structures that use a high Purcell effect to enhance the defects locally, the reflective dielectric cavity mainly uses a reflective metal layer and dielectric layer to improve the collection efficiency. Specifically, the transmission dissipation loss is reduced, and the metal reflective layer increases the upward fluorescence and is more directional. Reflection of the excitation laser by the metal reflection layer can play the role of multiple repumping. In conclusion, the reflective dielectric cavity exhibits good fluorescence and ODMR enhancement.

\section*{Method}

\textbf{Sample preparation.}
S1813 photoresist was first spin-coated on a clean SiO2/Si substrate at a speed of 4000 rpm for 40 s and baked at 115 °C for 1 min. Then, we used standard processes of ultraviolet lithography (SUSS MABA6) and developed for 50 s to fabricate the coplanar waveguide pattern. Subsequently, we coated 120-nm aluminum film and $\rm SiO_2$ dielectric layers of different thicknesses (0 nm, 20 nm, 50 nm, 80 nm, 110 nm, 140 nm) by an LAB18 E-Beam Evaporator to form the structure of RDC. Next, we transferred the hBN flake onto the RDC. Finally, we created a  $\rm V_B^-$ spin defect by helium focused ion beam system with a $\rm 10^{17}$-ions/$\rm cm^2$ implantation dose and 30-keV implantation energy.

\textbf{Simulations.}
All numerical simulation results are obtained using the finite difference time domain (FDTD) method, including  the Purcell factor, enhancement factor, and propagation time slice diagram of the light field. The refractive index of hBN is set to 2.0, and the rest use data from the material library. The minimum mesh size is set to 0.25 $\mu$m.

\textbf{Experimental setup.}
We used a home-made confocal microscope with a 0.9-N.A. objective (Olympus MPLFLN100x) to measure the hBN sample. The fluorescence was excited by 532-nm laser, collected by the same  objective and then reflected into the collection device through a beam splitter (BSW26R, Thorlabs) and a 750-nm long-pass filter (FELH0750 Thorlabs) to filter the laser and background fluorescence. The filtered signal light was collected using a single-mode fiber, and the fluorescence counts were detected by a silicon-based avalanche diode. For the ODMR measurement, we employed an aluminum waveguide on the substrate to transmit the microwave and used a microwave field signal generator to generate the microwave (SSG-6000RC, Mini-Circuits) and a power amplifier (ZHL-20W-13SW+, Mini-Circuits) to enhance the microwave power. In addition, we used a white light picosecond pulsed laser (SuperK Ehtreme, NKT Photonics) and PicoQuant for fluorescence lifetime measurement (HydraHarp).

\section*{Acknowledgments}

This work is supported by the Innovation Program for Quantum Science and Technology (No. 2021ZD0301200), the National Natural Science Foundation of China (Nos. 12174370, 12174376, and 11821404), the Youth Innovation Promotion Association of Chinese Academy of Sciences (No. 2017492), the Open Research Projects of Zhejiang Lab (No.2021MB0AB02), the Fok Ying-Tong Education Foundation (No. 171007). This work was partially carried out at the USTC Center for Micro and Nanoscale Research and Fabrication.

%\bibliographystyle{acs.bst}
%\bibliography{ref.bib}

\end{document}